# Statistical position reconstruction for RPC-based thermal neutron detectors

A. Morozov[1,*], L.M.S. Margato[1], V. Solovov[1], A. Blanco[1], J. Saraiva[1], T. Wilpert[2], K. Zeitelhack[3], K. Roemer[4], C. Höglund[5,6], L. Robinson[5], R. Hall-Wilton[5,7]

[1]LIP-Coimbra, Departamento de Física, Universidade de Coimbra, Rua Larga, 3004-516 Coimbra, Portugal

[2]Helmholtz-Zentrum Berlin für Materialien und Energie, Hahn-Meitner-Platz 1, 14109 Berlin, Germany

[3]Heinz Maier-Leibnitz Zentrum (MLZ), FRM-II, Technische Universität München, D-85748 Garching, Germany

[4]Helmholtz-Zentrum Dresden-Rossendorf, Bautzner Landstraße 400, 01328 Dresden, Germany

[5]European Spallation Source ERIC (ESS), P.O Box 176, SE-221 00 Lund, Sweden

[6]Impact Coatings AB, Westmansgatan 29G, SE-582 16 Linköping, Sweden

[7]University of Milano-Bicocca, Department of Physics, Piazza della Scienza 3, 20126 Milan, Italy

[*]Email: andrei@coimbra.lip.pt

## Abstract

Multilayer position-sensitive $^{10}$B-RPC thermal neutron detectors offer an attractive combination of sub-millimeter spatial resolution and high (>50%) detection efficiency. Here we describe a new position reconstruction method based on a statistical approach. Using experimental data, we compare the performance of this method with that of the centroid reconstruction. Both methods result in a similar image linearity/uniformity and spatial resolution. However, the statistical method allows to improve the image quality at the detector periphery, offers more flexible event filtering and allows to develop automatic quality monitoring procedures for early detection of situations when a change in the detector operation conditions starts to affect reconstruction quality.

## 1. Introduction

The concept of $^{10}$B-RPC position-sensitive thermal neutron detector [1] combines two technologies: resistive plate chambers (RPCs) [2] and $B_4C$ thermal neutron converters [3] deposited at the metallic cathodes of hybrid RPCs. It has already been shown that detectors based on this concept can simultaneously offer high detection efficiency (~50%) and spatial resolution in two dimensions of about 0.25 mm FWHM (full width at half maximum) [1,4].

A hybrid double-gap RPC with a metallic cathode, covered on both sides with the converter, is an elementary building block of $^{10}$B-RPCs. The cathode is positioned between two resistive anodes, forming two gas-gaps [1]. The position sensitivity is provided by installing two arrays of orthogonal signal pick-up strips close to each anode. One $^{10}$B-RPC has only ~5% neutron detection efficiency. However, this value can be increased to ~50% by constructing the detector with a stack of such $^{10}$B-RPCs blocks [1,5]. In this case the arrays of signal pick-up strips can be shared by two neighboring gas-gaps [4] and the cathode signals can be used to identify the triggered RPC.

All previous studies dedicated to $^{10}$B-RPC detectors have utilized centroid-based position reconstruction [4,5]. This is a very straightforward method which, in order to reconstruct a neutron capture event, only requires the knowledge of the charge induced in the strips and the strip positions. The centroid approach has intrinsic drawbacks, such as distortions appearing at the periphery of the detector, intrinsic non-linearity and a limited capability to discriminates dark count events.

In this study we describe a new statistical reconstruction approach based on a mathematical model of the spatial dependence of the response of the signal pick-up strips. The response parameterisation is validated using experimental data recorded with a $^{10}$B-RPC detector prototype at the V17 monochromatic neutron beamline (3.35 Å) of the Helmholtz-Zentrum Berlin (HZB). We also give a detailed comparison of the performance of the statistical and the centroid reconstruction in terms of the image quality and spatial resolution.

## 2. Methods

### 2.1 Detector

This study was performed with a $^{10}$B-RPC detector (see figure 1) with one hybrid double-gap RPC [1]. The aluminium cathode (0.5 mm thick, 100×100 mm$^2$) is covered on both sides with a 1.15 μm thick layer of sputtered B$_4$C with enrichment of 97% $^{10}$B. The deposition was conducted at the ESS Detector Coatings Workshop in Linköping [3,6,7]. Both RPCs sharing the cathode have the same configuration. The 0.35 mm wide gas-gaps are defined by the diameter of the spacers made of nylon monofilaments (there are two spacers per gas-gap with a separation of 250 mm). The anodes are made from 0.5 mm thick float glass. On the side opposite to the gas-gap, the anodes are covered with a ~0.05 mm layer of ink (1·10$^8$ Ω/□ surface resistivity) to uniformly distribute the electric potential over the electrode area. A negative potential of 2300 V is applied to the cathode, and the anodes are at the ground potential. The detector is filled with R134A (C$_2$H$_2$F$_4$) at atmospheric pressure maintaining a gas flow of about 2 cm$^3$/minute.

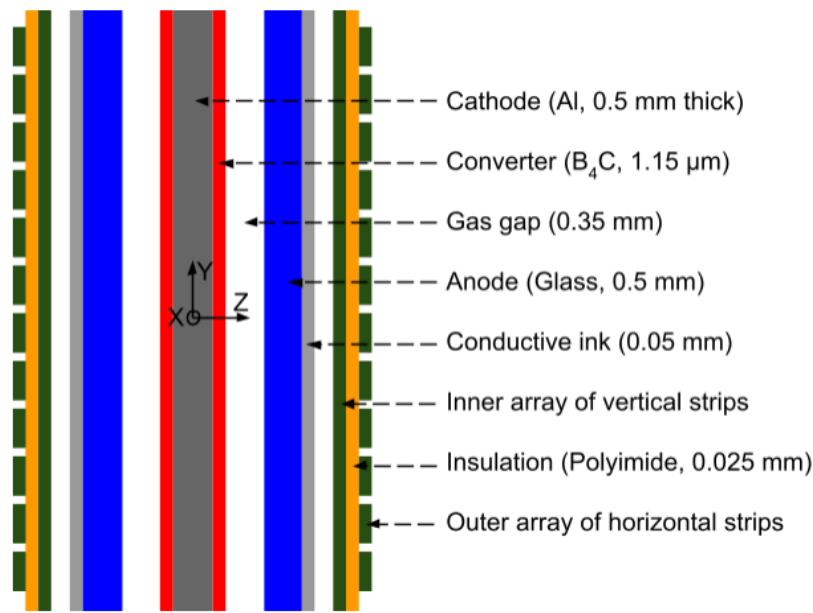

Figure 1. Cross-section of the $^{10}$B-RPC detector with one double-gap RPCs. The left and right sides of the detector are symmetric. Relative sizes are not to scale. The inner and outer arrays are used to readout X and Y coordinate, respectively. The signal pick-up strips of the inner arrays on the left and right sides of the detector, situated at the same X position are interconnected and fed to the same ADC channel. Similarly, the strips of the outer arrays situated at the same Y position are also interconnected.

To organize the signal readout, a flexible printed circuit board (PCB) containing two arrays of signal pick-up strips is installed in front of each anode. The PCBs have two copper layers (0.018 mm thick) insulated by a 0.025 mm thick polyimide film. The copper layers are etched to shape two mutually-orthogonal arrays, both with the pitch of 1 mm. The array facing the anode (the *inner* array) has 0.3 mm wide strips, thus leaving enough space between them to avoid excessive screening of the strips of the second array (the *outer* array). The outer array has 0.9 mm wide strips.

The inner and the outer arrays are used to read orthogonal coordinates X and Y on the detector plane of the neutron capture positions. Since the arrays are identical for both sides of the double-gap RPC, and a neutron event is localized to a single gas-gap, the strips at the same X (and, similarly, the same Y) coordinate for both RPCs are connected to the same readout channel.

The readout of the signals from the strips and the cathode is performed by charge-sensitive preamplifiers connected to a 40-MHz 10-bit waveform digitizer based on the GSI-developed TRB board with two ADC add-ons [8]. The signals from 43 strips in each directions are collected, resulting in a 42×42 $mm^2$ readout area of the detector. Event triggering is performed using the cathode signal. For each event, the signal waveforms with 80 samples are collected, which corresponds to 20 µs duration (downsampling of 10 is used). A part of the waveform is recorded before the trigger, allowing to establish the baseline. The signal processing approach adopted in this study is discussed in section 2.3. The relative gain factors of all electronic channels were confirmed to be equal within ±2% using a calibrated pulser to feed one channel at a time with the same charge.

## 2.2 Setup

All experimental data were collected at the V17 beam station at the HZB using a monochromatic neutron beam (3.35 Å). The detector was installed in an aluminium enclosure with a 1 mm thick entrance window. The beam, irradiating the detector normally, was collimated to have the shape approximately equal to the readout area of the detector. The beam was attenuated with boron-containing glass plates and it was confirmed that the counting rate of the detector scales linearly with the neutron flux [9].

Several masks were used to shape the beam. One of them was a 1 mm thick cadmium plate with a 0.2 mm wide and 23 mm long slit. The mask was installed directly in front of the detector (<10 mm from the RPC) on a remotely-controlled high-precision XY table. Using this arrangement we recorded datasets with the position of the slit scanning the field of view of the detector with a 0.1 mm step in the direction perpendicular to that of the slit (X for the vertical and Y for the horizontal slit orientation).

Two more masks with the grooves forming words "LIP FRM II" (1 mm thick cadmium) and "HZB" (0.25 mm thick gadolinium) were used to record the data used to assess linearity and uniformity of the reconstructed images. These masks were installed directly at the surface of the entrance window of the detector (<5 mm from the RPC).

## 2.3 Signal processing

An example of a signal waveform recorded for one of the strips is shown in figure 2 (left).

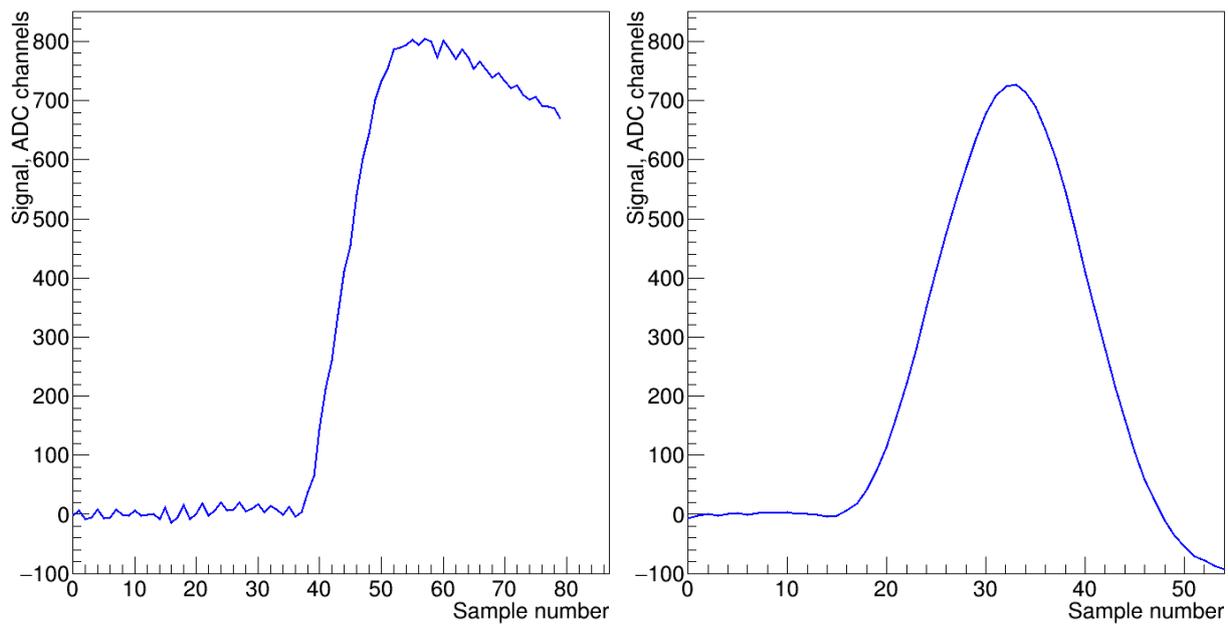

Figure 2. Signal waveform (left) and the same waveform post-processed using the trapezoidal filter (right). The time interval between the samples is 250 ns.

In order to reduce high frequency noise and electromagnetic interference, the signal waveforms were digitally post-processed using the trapezoidal filter [10] with both constants of 8 samples (see figure 2, right). The length of the waveform region digitized before the signal onset allows to correct for the offsets in the baseline. The amplitude of the peak of the processed waveform is considered to be the signal amplitude. If the processed peak appears at the sample indices less than 25 or larger than 36 (waveform with very low signals or distorted waveforms), the signal amplitude is considered to be zero.

An example of the signal amplitude distribution across the strips in X (blue) and Y (red) directions for a neutron event is shown in figure 3. The strip indices are shifted independently for X and Y directions for convenience of presentation, so to obtain the two largest amplitudes on indices 0 and 1. The remaining strips have zero amplitude.

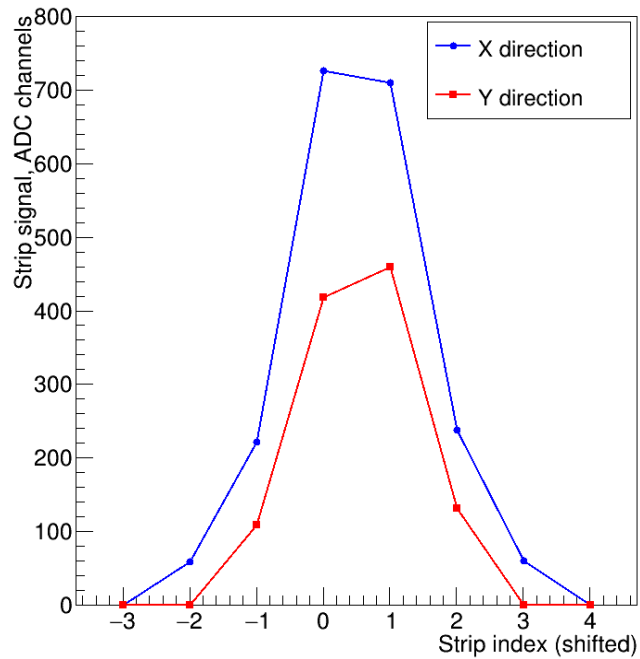

Figure 3. Example of the distribution of the strip signal amplitudes vs shifted strip index for a neutron event. All other strips have zero signal. Strip pitch is 1 mm for both directions.

The cathode signals have a very similar shape to those from the signal pick-up strips. Therefore, the same trapezoid filtering was applied to them as well. The distribution of the cathode signal amplitudes for a dataset recorded with flood field irradiation is shown in figure 4 (left). Note the presence of saturation (the peak at ~4400).

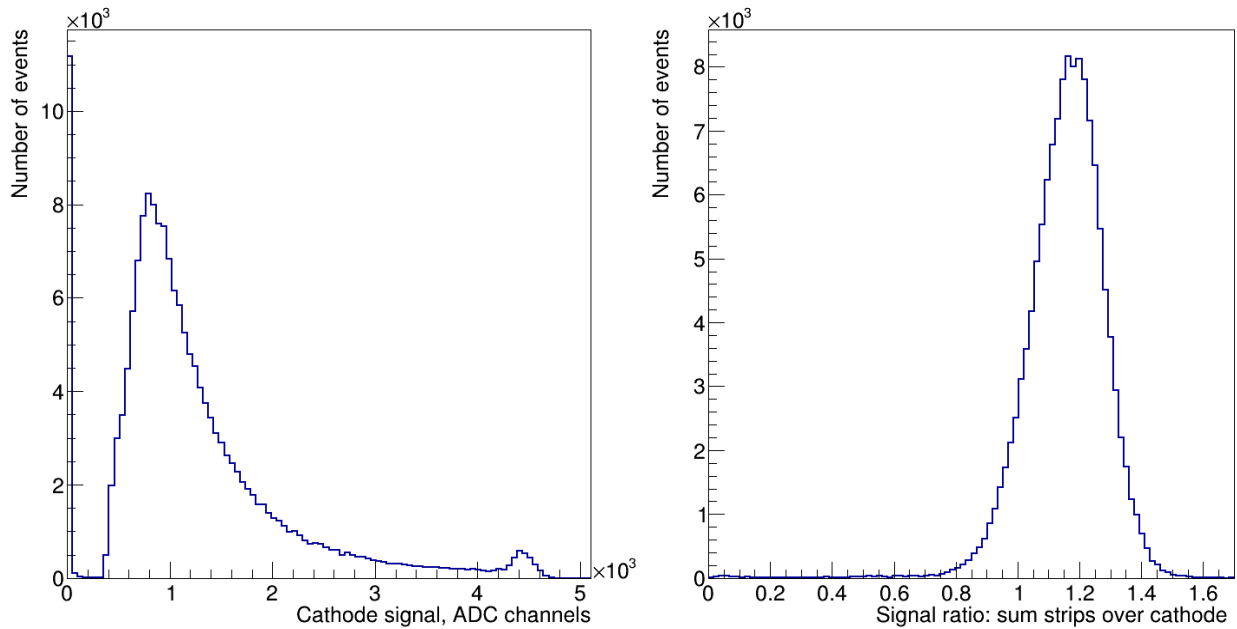

Figure 4. Left: Distribution of the cathode signal amplitudes obtained for a dataset recorded with flood-field irradiation. Right: Distribution of the amplitude ratios of the sum signal in all strips (X and Y directions combined) and the cathode signal.

The distribution of the amplitude ratios of sum signals in all strips to the cathode signal is shown in figure 4 (right). The fact that the cathode signal is smaller than the sum strip signal is explained by the fact that the output of the cathode preamplifier is connected both to the trigger circuit and to one of the ADC channels, and so the signal is split (one-to-one ratio).

## 2.4 Centroid reconstruction of event positions

Centroid reconstruction was performed independently for X and Y directions using the following procedure:

1) Find the strip $i_{max}$ with the maximum signal;

2) Select the strips with the indices from $i_{max}$-N to $i_{max}$+N with the signal above a given threshold (if the event is close to the periphery the total number of selected strip can be less than 2N+1);

3) If the number of selected strips is less than 2, ignore the event;

4) Over the selected strips compute the sum of the strip signals and the sum of the strip signal multiplied by the strip center coordinate;

5) The event position is given by the ratio of the latter and the former sums.

The best image quality was obtained with the threshold of about 20 ADC channels and the values of N from 4 to 6. It was also found that several event filtering procedures improve the reconstruction quality by eliminating "ghost" features (probably triggered by the dark counts) and slightly improving the spatial resolution and linearity.

An event was suppressed if:

a) The cathode signal is below 300 or above 4200 ADC channels;

b) The ratio of the sum of the strip signals (X and Y combined) and the cathode signal is below 0.8 or above 1.5;

c) The ratio of the strip sum signal in X direction to that in Y directions is below 0.8 or above 2.8.

Using datasets recorded with a narrow slit, the numeric values appearing in the criteria (b) and (c) were chosen to provide an effective filter suppressing the ghost features, and, in the same time, removing only a negligible number of events in the irradiated area.

A minor improvement in the uniformity was also achieved by subtracting the threshold level from signals of all the strips before the sum calculation (step 4 above). It was confirmed that this correction does not affect the spatial resolution, most likely since the threshold is ~50 times smaller than the strip sum signal value of the weakest event.

## 2.5 Strip response function

In order to perform statistical reconstruction, a model has to be defined which describes the dependence of the strip signal on the lateral distance (projection on the RPC plane) between the strip center and the event position. Here we designate this dependence as the *strip response function* (SRF).

SRFs of several vertical and horizontal strips were measured by scanning the active area of the detector with a vertically/horizontally oriented 0.2 mm wide slit with a 0.1 mm step. Note that this step is small compared to the FWHM of the expected strip response of ~2 mm (see figure 3). Since the total amount of the induced charge fluctuates quite a lot from event to event (figure 4, left), the SRF should describe the signal ratio of the strip and the cathode. The cathode signal is thus used as a measure of the total charge induced in a particular event.

A procedure was defined which allows to compute the average signal for each strip for a dataset recorded at a particular slit position. The first step is to filter out the background events. This is achieved by analyzing the results of the centroid reconstruction and introducing a spatial filter on the event positions by defining a narrow rectangle (~1 mm wide) around the center of the slit image. As an example, figure 5 shows a histogram of the signal ratio of a specific strip and the cathode for all events from a dataset with the spatial filter already applied. Since the total number of events ($3\times10^4$) recorded at each slit position is relatively low and some background events still leak through the spatial filter, the average value of the signal of a strip is strongly affected by statistical fluctuations and thus is not a good estimate of the strip response. We have chosen to use instead the mean value given by the Gaussian fit of a histogram with the strip-to-cathode signal amplitude ratio (figure 5).

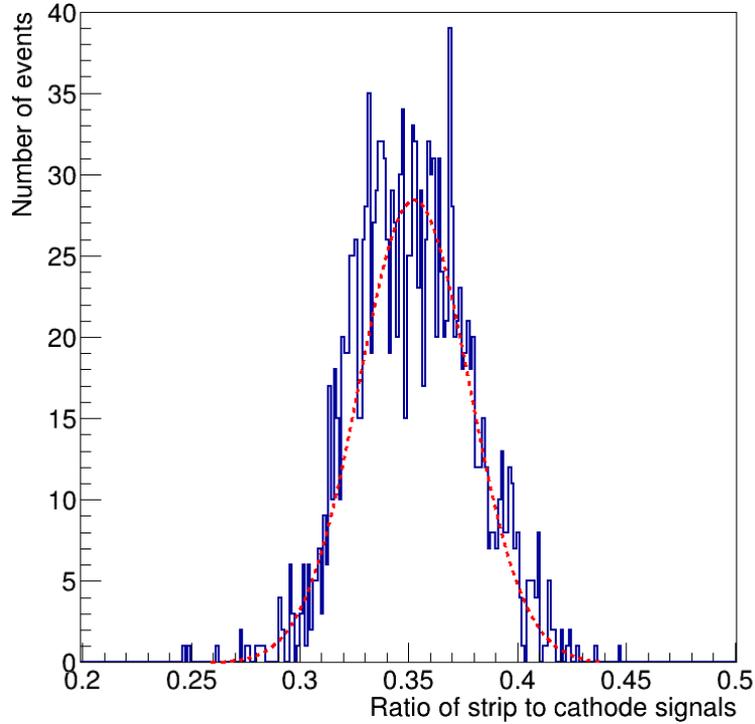

Figure 5. Histogram of the strip-to-cathode signal amplitude ratio for the events from a filtered dataset recorded at a given slit position. The red dashed curve shows a Gaussian fit.

Combining the data from the datasets acquired at different slit positions with respect to a particular strip it is possible to obtain the SRF curve. An example of such curve is given in figure 6 (blue dots).

According to the analytical model of the RPC strip response [11,12], for the RPC geometry used in this study (anode thickness is similar to that of the gas-gap and smaller than the strip pitch) the SRF of the strip with index $i$ can be parameterized in the form:

$$SRF_i(x) = \frac{A}{cosh(W(x-x_i))} \qquad (1)$$

where $x_i$ is the position of the strip center, $x$ is the event position, $A$ is a scaling factor and $W$ is a parameter characterizing the width of the response profile. It was indeed possible to make a very good fit using this expression for all experimental SRF data obtained in this study. An example of the fit result is shown in figure 6 (red solid line).

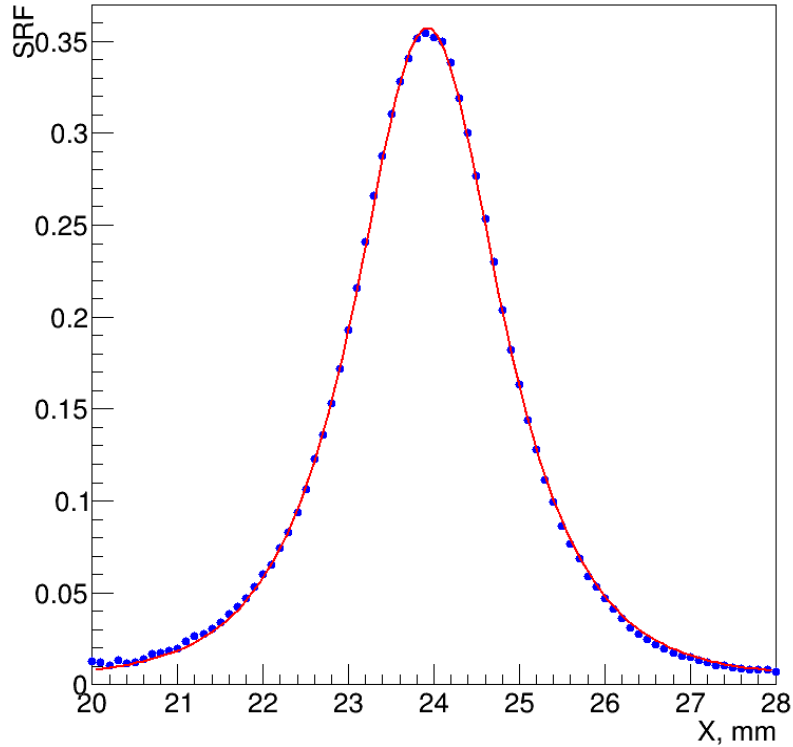

Figure 6. SRF data obtained experimentally (blue dots) and the result of a fit according to equation (1).

## 2.6 Statistical reconstruction of event positions

The statistical approach for position reconstruction is based on finding the event position which gives the best possible match between the observed strip signals and the corresponding signals predicted using the SRFs. Statistical reconstruction methods were developed for 2D-sensitive Anger cameras [13] and are known to be capable to provide better image linearity than the centroid method [14].

The statistical reconstruction is performed independently for X and Y directions assuming that the SRFs of all strips are well parameterized by equation 1. For each event the strip with the strongest signal and three strips on each side (*active* strips, 7 in total or less if the event is close to the periphery) are taken into account. The profiles with these strip signals are fitted according to the equation 1 using the minimizer from CERN ROOT toolkit [15, 16] (MIGRAD algorithm) and the obtained $x$ values give the positions of the events. Since the measured SRF profiles show that the value of W strongly fluctuates over the detector sensitive area (see discussion in section 3.1), the fitting was performed over all three parameters ($A$, $W$ and $x$). The initial values of $x$, $A$ and $W$ were set to the position of the strip with the strongest signal, 1 and 1.5, respectively.

In contrast to the methods described in [14], this method does not require exact knowledge of the SRF parameters, it just assumes that the parameterization according to equation 1 is valid for all possible event positions. This is an important property as the experimental results given below show that the width of the SRF of a given strip can change over the active area of the detector due to, for example, non-uniformity of the gas-gap thickness.

Several event filtering procedures can be applied to improve the image quality. The event was rejected if:

a) The cathode signal is below 300 or above 4200 ADC channels (triggered by noise or saturated event);

b) The value of W given by the fit is unrealistically low (less than 1.0);

c) The value of $\chi^2$ for X or Y direction is above a certain threshold, indicating an error in the reconstruction or a distorted event:

$$\chi^2 = \sum_i \frac{(S_i - E_i)^2}{E_i} \qquad (2)$$

where $S_i$ and $E_i$ are the measured and the expected (based on the SRFs) signals in the i-th strip and the sum is performed over the active strips. The $\chi^2$ filtering threshold is discussed below.

## 3. Results

### 3.1 Comparison of the centroid and statistical reconstruction

The heatmaps of the reconstructed event density vs XY event position for a dataset recorded with flood-field irradiation are shown in figure 7 for the centroid (left) and statistical (right) reconstruction methods. The entire readout area of the detector (42×42 mm$^2$) is shown with the bin size of 0.2 mm.

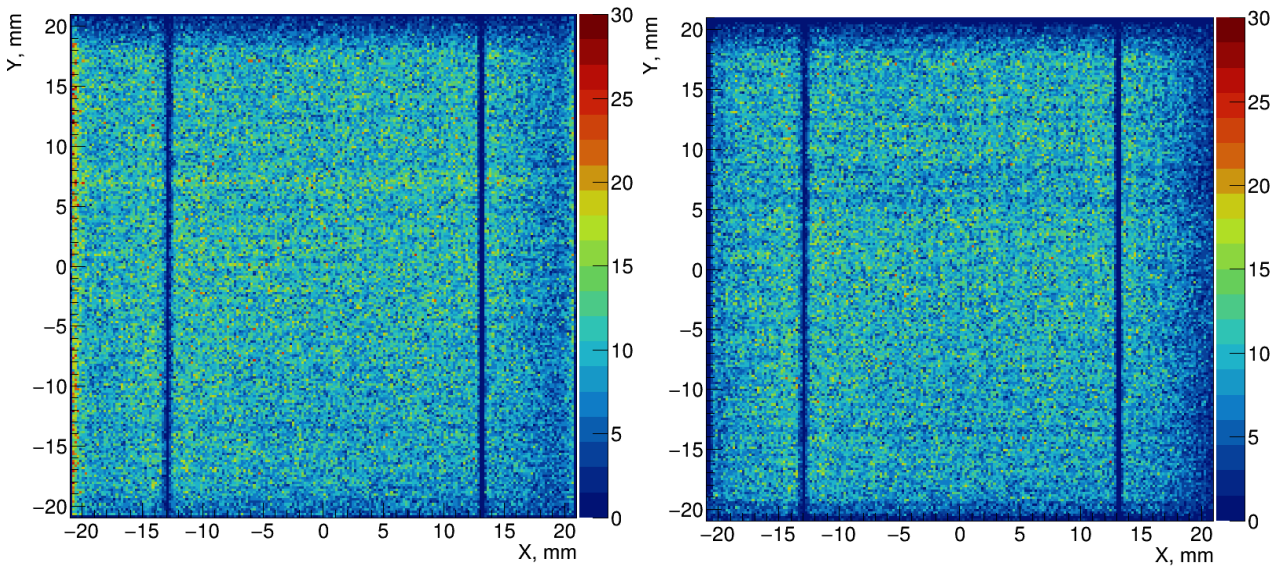

Figure 7. Heatmaps of the reconstructed event density (number of events per bin) for a dataset recorded with flood field irradiation: centroid (left) and statistical (right) methods.

The dark vertical lines are the artifacts appearing due to the presence of the spacers in the gas-gaps. The significant non-uniformity at the left edge of the heatmap obtained with the centroid reconstruction (figure 7, left) is explained by two factors: (1) the irradiated area extends to the left side more than the size of the readout area of the detector, and (2) the centroid reconstruction has a systematic distortion due to the fact that the events at the periphery are reconstructed using the signals from only a fraction (down to a half) of all the strips with the induced charge. This pulls the coordinates of the reconstructed peripheral events in the direction of the detector center, creating artificially higher-density region clearly visible at the left edge of the plot (X ranging from -21 to -20 mm).

Both heatmaps also show significant density fluctuations in a ~2 mm wide horizontal area at Y of 7 mm. A possible explanation of this feature is given in section 3.4. The origin of two thin horizontal dark lines at Y of ±13 mm appearing for both reconstruction methods has not been identified.

The distributions of *W* and *A* fit parameters as well as the heatmap of $\chi^2$ obtained during statistical reconstruction (Y direction) of this dataset are given in figures 8 and 9. The $\chi^2$ heatmap has elevated values in the same area (Y from ~6 to ~8 mm) where the heatmap of the reconstructed density exhibits strong non-uniformity. Analysis of the signals also show that the strip situated at Y of 7 mm has more than double contribution to the $\chi^2$ values compared to any other strips, suggesting that there is a problem with the signal read-out from this strip. As already mentioned above, further discussion of this image distortion is given in section 3.4.

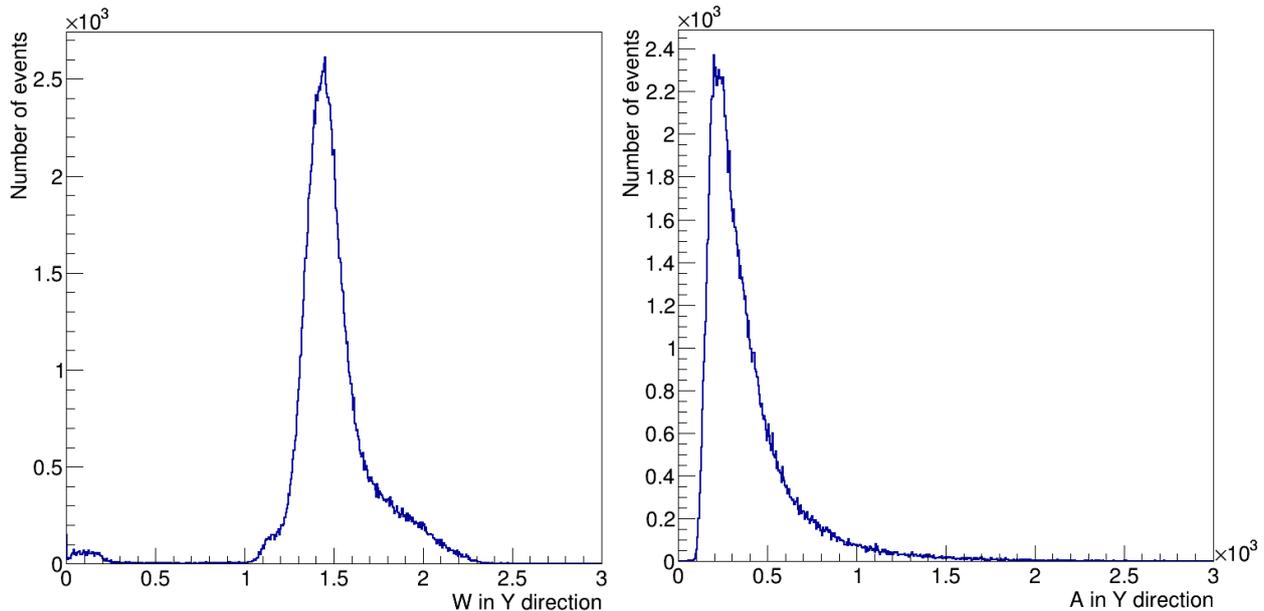

Figure 8. Distributions of the *W* (left) and *A* (right) parameter values obtained during Y coordinate reconstruction (flood irradiation dataset).

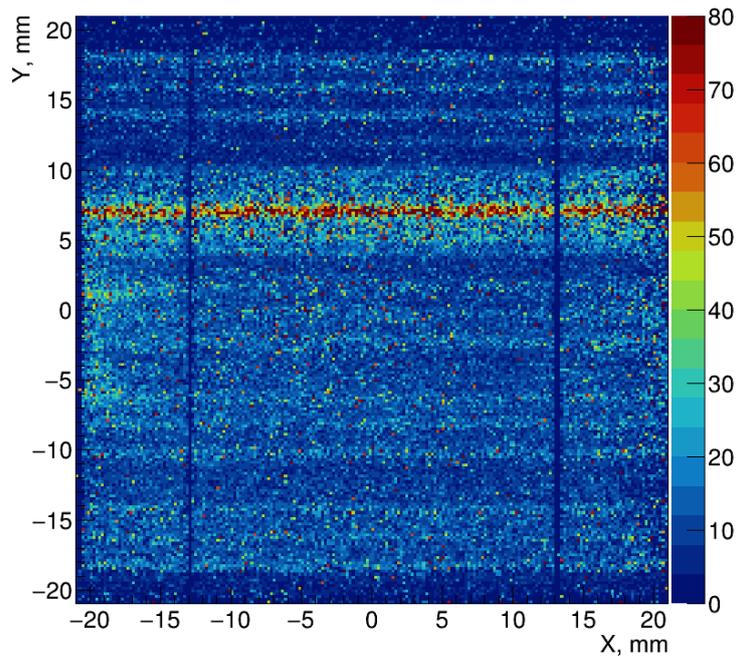

Figure 9. Heatmap of $\chi^2$ (average value per bin) for Y coordinate reconstruction as a function of the reconstructed position. Dark-red color (maximum of the color-coded scale) indicates bins with $\chi^2$ values equal or larger than the value used as the filter threshold during the statistical reconstruction.

Excluding the periphery (2 mm from the edges) where the centroid reconstruction is strongly affected by the non-linearity, the number of accepted events for statistical reconstruction is 6% less compared to the one obtained with the centroid. This result indicates that the effective detection efficiency is negatively affected, however, the applied threshold for the $\chi^2$ filter in Y direction might be too restrictive as a result of the effort to suppress the artifact appearing at Y of 7 mm (see figure 9), which leads to rejection of a fraction of good events.

Figures 10 and 11 show examples of the heatmaps reconstructed for the datasets recorded with the masks with engraved letters. The grove widths for the letters are 1 mm for the first and 0.4 mm for the second mask. The image bin size is 0.1 mm in both cases. The heatmaps show a high degree of linearity and uniformity. The only area with noticeable distortions appears again at Y of about 7 mm: the "Z" letter has the lower horizontal segment slightly wider than the upper.

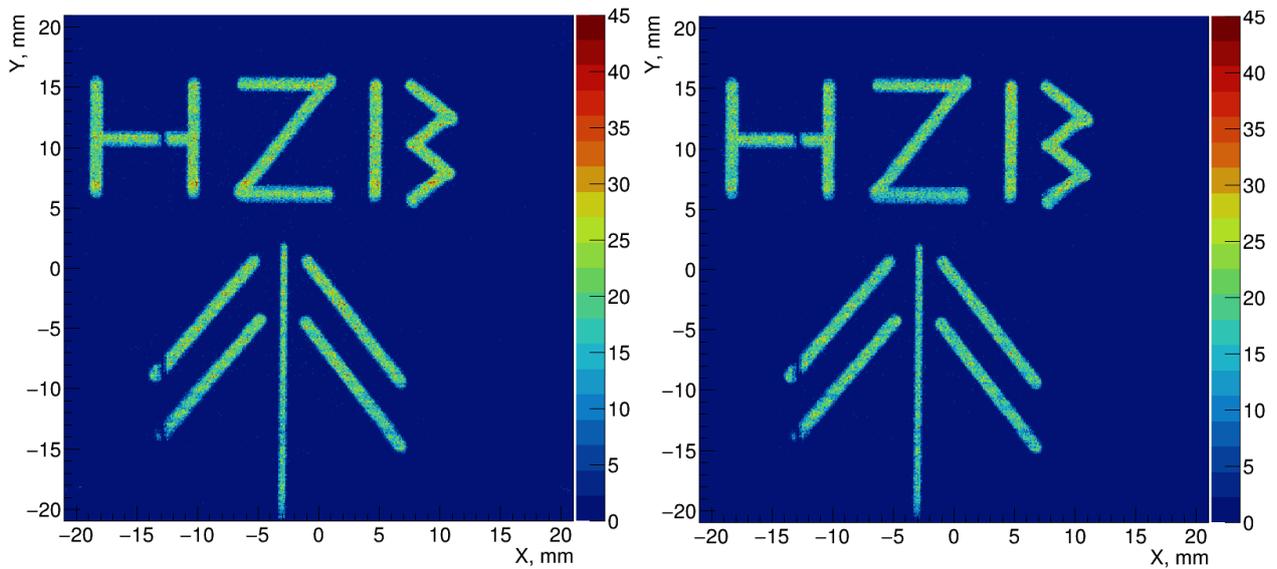

Figure 10. Heatmaps of the reconstructed event density (number of events per bin) obtained using the centroid (left) and the statistical (right) methods for a dataset recorded with a 0.25 mm thick gadolinium "HZB" mask.

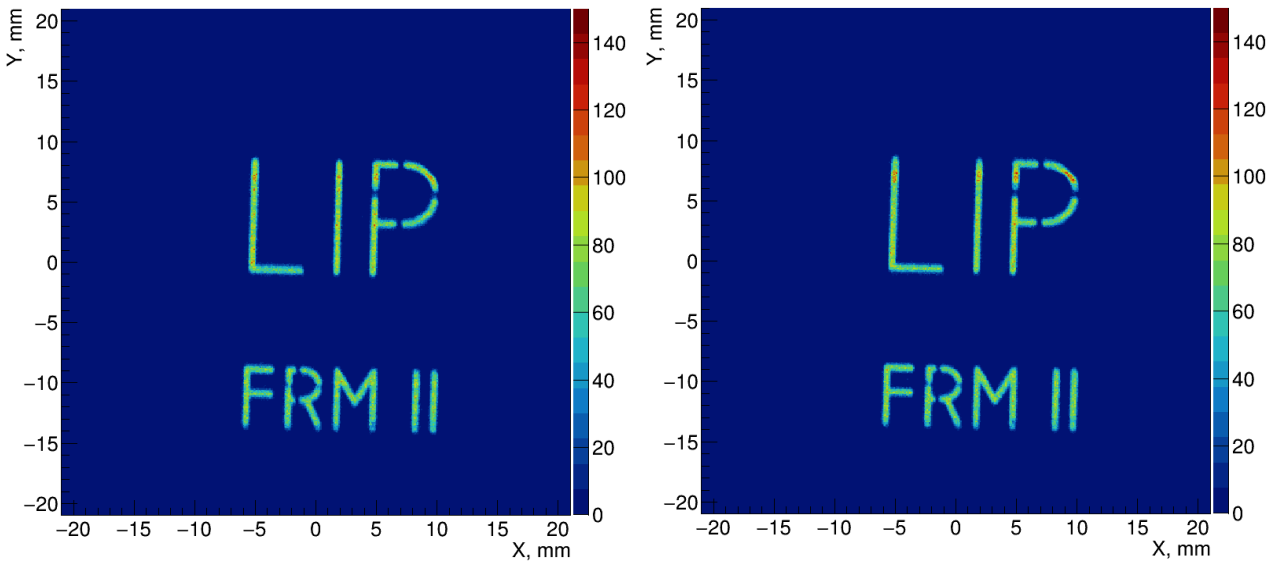

Figure 11. Heatmaps of the reconstructed event density (number of events per bin) obtained using the centroid (left) and the statistical (right) methods for a dataset recorded with a 1 mm thick cadmium "LIP FRM II" mask.

The datasets recorded with a single 0.2 mm wide slit were used to evaluate the difference in the spatial resolution obtained with the two reconstruction methods. Note that this study was not dedicated (and not equipped) to accurately evaluate the spatial resolution of the detector: the narrowest slit available at the time of the measurements has the width comparable with the resolution. Also, while all possible measures were taken to minimize the beam divergence, it was not accurately measured.

The centroid-reconstructed images obtained with the slit positioned vertically and horizontally, as well as the projections along the slit direction are shown in figures 12 and 13. Application of the statistical reconstruction method results in essentially the same images and profiles. Since the slit was positioned with a small angle in respect to the vertical (or horizontal) direction, the projections were obtained taking this angle into account, and thus the coordinates of the profiles are given as a function of the distance from the center of the selected tilted area.

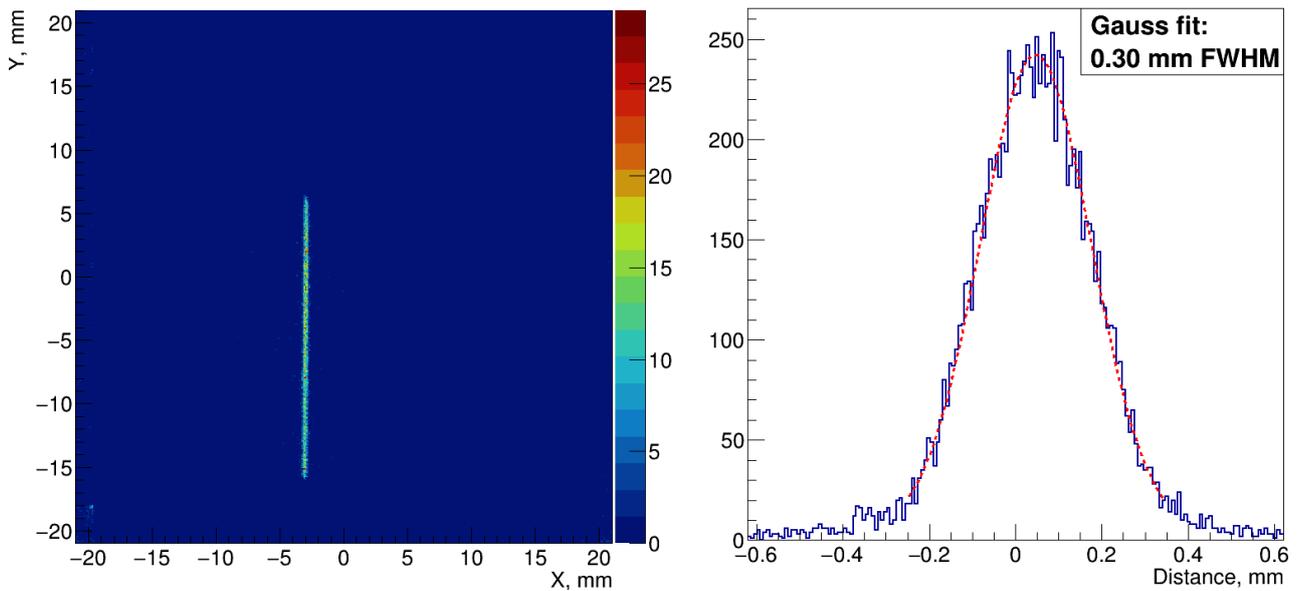

Figure 12. Left: Heatmap of the reconstructed event density (number of events per bin) obtained with the centroid method for a dataset recorded with a vertical slit. Right: Projection of the heatmap along the slit direction. Red dashed line shows a Gaussian fit.

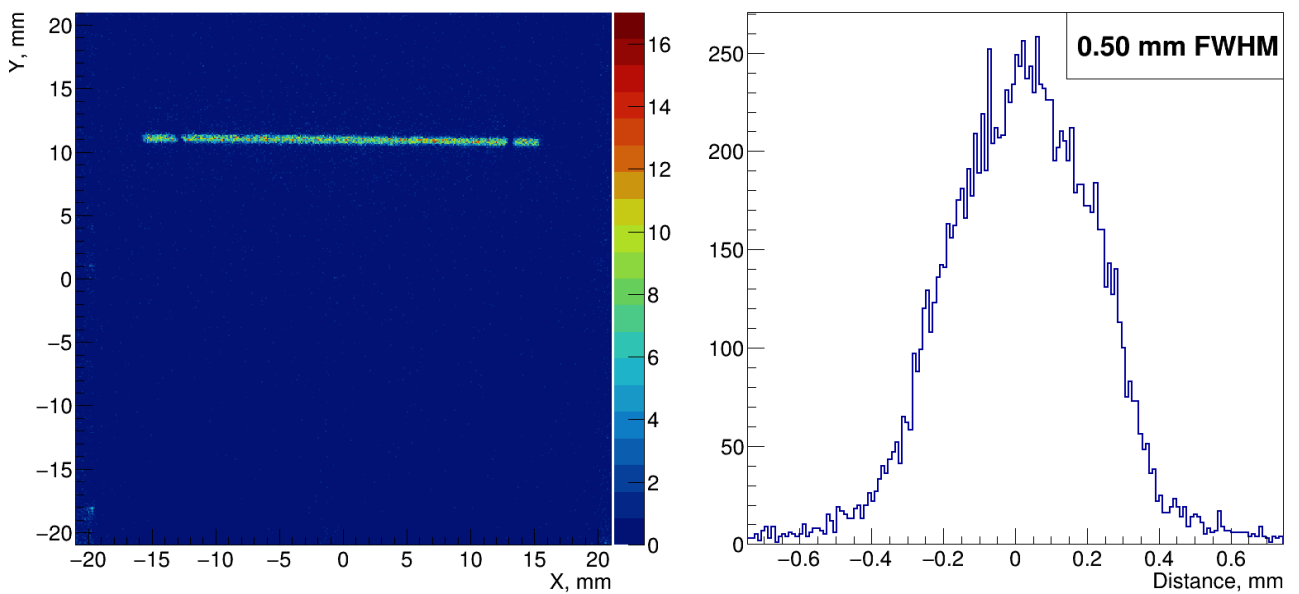

Figure 13. Left: Heatmap of the reconstructed event density (number of events per bin) obtained with the centroid method for a dataset recorded with a horizontal slit. Right: Projection of the heatmap along the slit direction. The shape of the projection is strongly non-Gaussian.

Figure 14 shows the mean (left) and FWHM (right) of the Gaussian fit of the profile for 41 datasets recorded with a vertical slit at X positions ranging from 1 to 5 mm with the step of 0.1 mm. The top and bottom rows give the centroid and the statistical reconstruction results, respectively. Note the good linearity in the reconstructed X position demonstrated by both methods: there is no modulation with the period equal to the 1 mm pitch of the strip arrays.

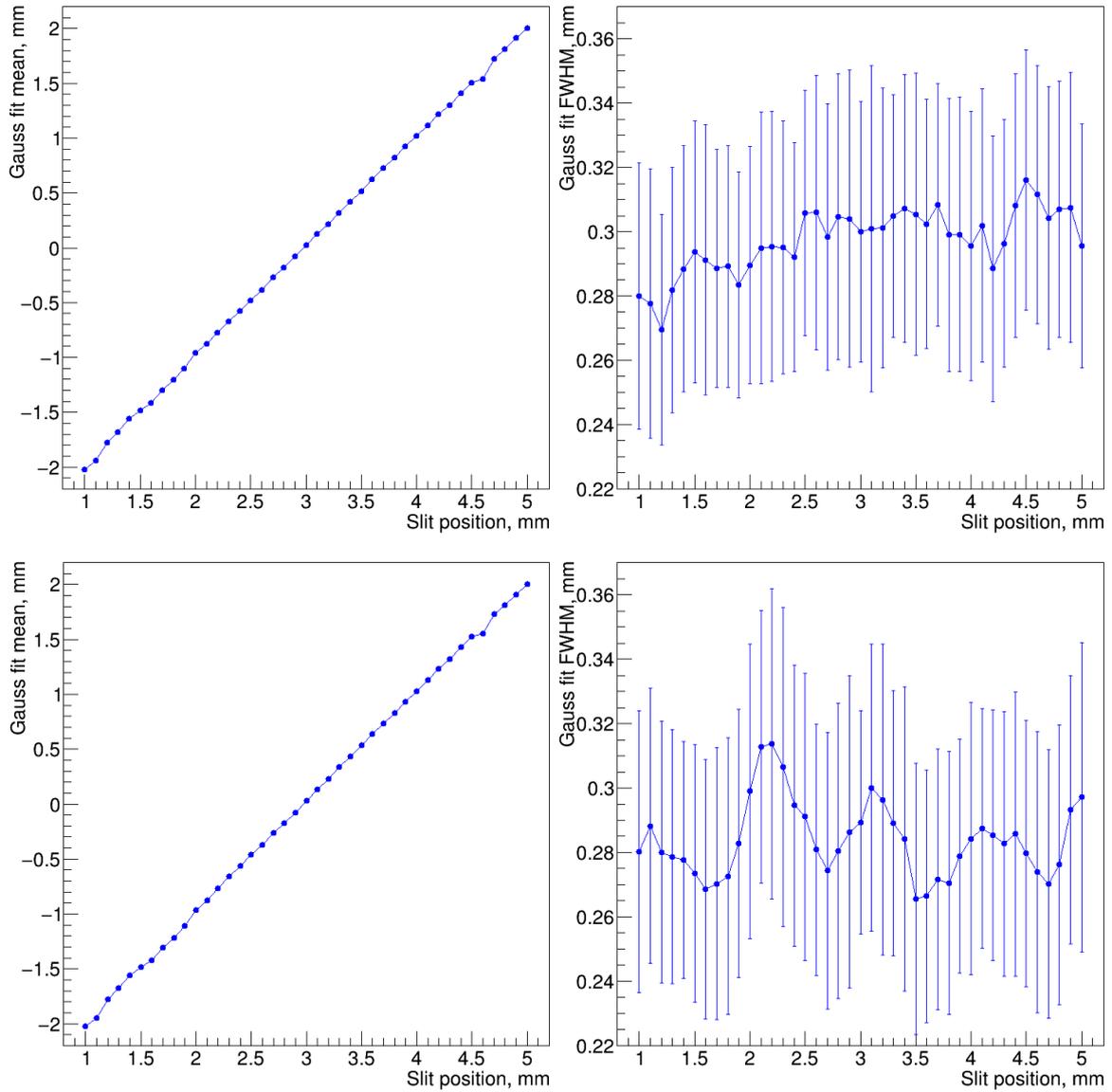

Figure 14. Mean (left) and FWHM (right) of the Gaussian fit of the profile for 41 datasets recorded with a vertical slit at X positions ranging from 1 to 5 mm with the step of 0.1 mm. The top and bottom rows give the centroid and statistical reconstruction results, respectively. The mean values are practically the same for the centroid and statistical reconstruction. The error bars show the uncertainties of the fit. For the mean, the uncertainties are smaller than the size of the markers.

The obtained results demonstrate that the spatial resolution in X direction for both methods is about 0.30 mm, and for Y direction is about 0.50 mm. The spatial resolution in Y direction is significantly worse (by ~65%) compared to that in X direction. The fact that the Y-direction array is situated behind the one in X direction in respect to the RPC anode cannot explain this effect since the induced signal amplitudes in both X and Y arrays differ only slightly: the average induced signal in Y direction is only 30% smaller than that in X direction. Also note that in Y direction the profiles are strongly non-Gaussian (see figure 13). A likely explanation of the difference in X and Y resolution is presented in the next section.

The measured values of the spatial resolution are influenced by multiple factors, most importantly the width of the slit, the divergence of the neutron beam, the average shift from the capture position for the avalanches formed in the gas-gap by the capture reaction products and the electronic noise. While this study is not equipped to measure and analyze the relative contributions of these factors, the statistical model of the strip response allows us to evaluate the contribution of the electronic noise in the digitized strip signals to the resolution.

We have generated a number of ideal events by using the equation 1, the experimental distributions of *A* and *W* parameters, and a set of fixed values of *x*. Then the signal in each channel was superimposed with a random value sampled according to the noise distribution, obtained from experimental data by considering only the strips situated far away (>15 mm) from the capture position. The distribution shows a range of signal amplitudes from 0 to 25 ADC channels with both mean and most probable value of about 8 (to be compared with the average signal amplitude at the strip response center of about 300). The constructed events were processed using the statistical and the centroid reconstruction methods. Figure 15 shows examples of the obtained distributions of the difference between the true and the reconstructed positions. All datasets generated with different *x* value gave essentially the same results. The widths of the obtained distributions show that the effect of the electronic noise on the uncertainty of the reconstructed position is about 0.017 mm for the statistical and 0.036 mm for the centroid methods, which is negligible with respect to the resolution of ~0.30 mm obtained with both methods in this study. However, these results suggest that the statistical method could result in a better ultimate spatial resolution compared to the centroid method in the conditions when the electronic noise is the dominating factor. Validation of this statement requires a dedicated experimental study attempting to reach those conditions, for example by using neutron beam with a very small divergence, a slit with a width on the order of 0.01 mm and an RPC with narrow gas gaps in order to reduce the lateral spread of the avalanches.

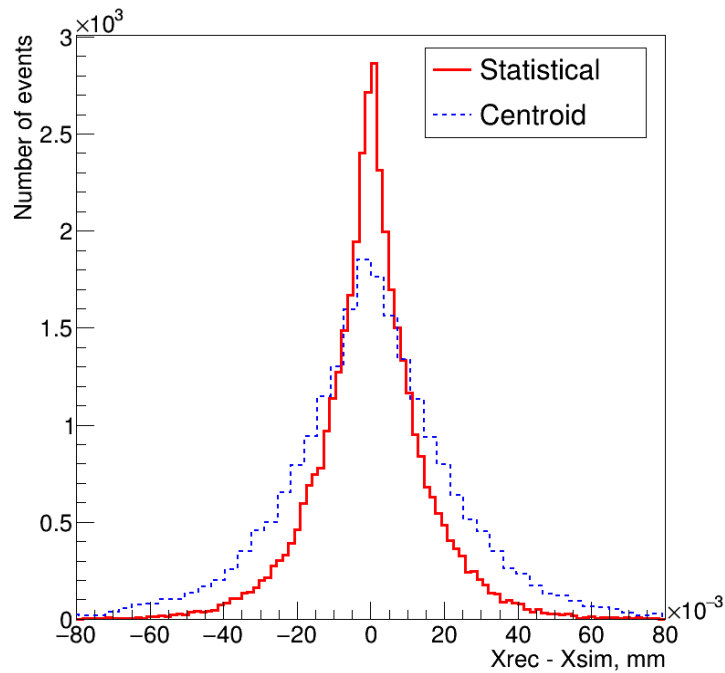

Figure 15. Distributions of the differences between the simulated and reconstructed event positions for the statistical and centroid methods. The data suggest that the contribution of the electronic noise to the uncertainty of the reconstructed position is about 0.02 mm and 0.04 mm for the statistical and the centroid methods, respectively.

## 3.2 Identification of the triggered gas-gap using the response width

A possible explanation of the difference in the spatial resolutions in X and Y directions is that there is a small (~0.2 mm) relative shift in the Y positions of the arrays with the horizontal strips between two sides of the double gap detector. In this case, the events triggered in different gas-gaps at the same Y coordinate would have a small shift in the reconstructed Y position. As described in section 2.1, the strips at the same Y position from both sides of the double-gap RPC are interconnected, so the events from the two gas-gaps are indistinguishable. Therefore, if this shift is smaller than the resolution of the detector, the reconstructed image of the 0.2 mm wide slit would be broadened by this effect resulting in worsening of the apparent resolution in this direction.

An analysis of the distribution of the *W* parameter (response width) of the fit shows that for several areas of the detector events are clearly divided in two groups with a different average value of *W* and approximately equal size. For example, a reconstructed image and the distribution of the *W* value for Y direction are given in figure 16 for a dataset recorded with a horizontal slit positioned at Y of 9 mm. One group has an average *W* value of 1.3 and the other a value of 1.5.

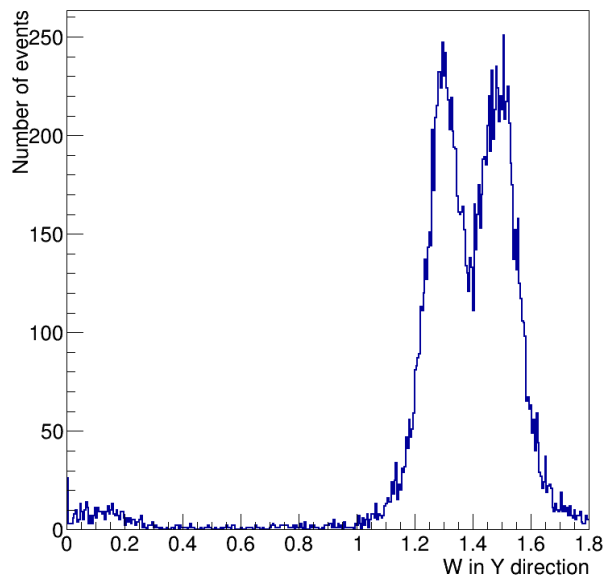

Figure 16. Distribution of *W* values for Y direction obtained for a dataset recorded with the horizontal slit positioned at Y of 9 mm.

The separation between the peaks is not constant over the field of view of the detector. This is apparent by comparing the distributions shown in figure 16 and the one obtained for the flood-field dataset (figure 8, left). Our results also show that the peak separation can change quite abruptly over the sensitive area of the detector: for example, a shift of 3 mm (from Y = 9 mm to Y = 12 mm) results in complete merging of the peaks. Note that a scan of the detector with a horizontal slit was performed only in the upper half of the detector. At the lower part just one measurement was taken with the slit positioned at Y of -7 mm. The distribution of W values for this dataset is essentially the same as the one obtained for Y = 9 mm (figure 16, right).

Assuming that the two peaks correspond to the events triggered in two different gas-gaps, the dataset recorded with the slit positioned at Y of 9 mm was divided in two using the value of W of 1.39 as the threshold. The reconstructed images for these two datasets indeed show narrow images of the slit with the Y center position separated by a distance of 0.21 mm. Figure 17 shows the projections obtained from both images along the slit direction together with their Gaussian fits. The width of the individual profiles is 0.31 mm FWHM, which is essentially the same as the spatial resolution obtained for X direction.

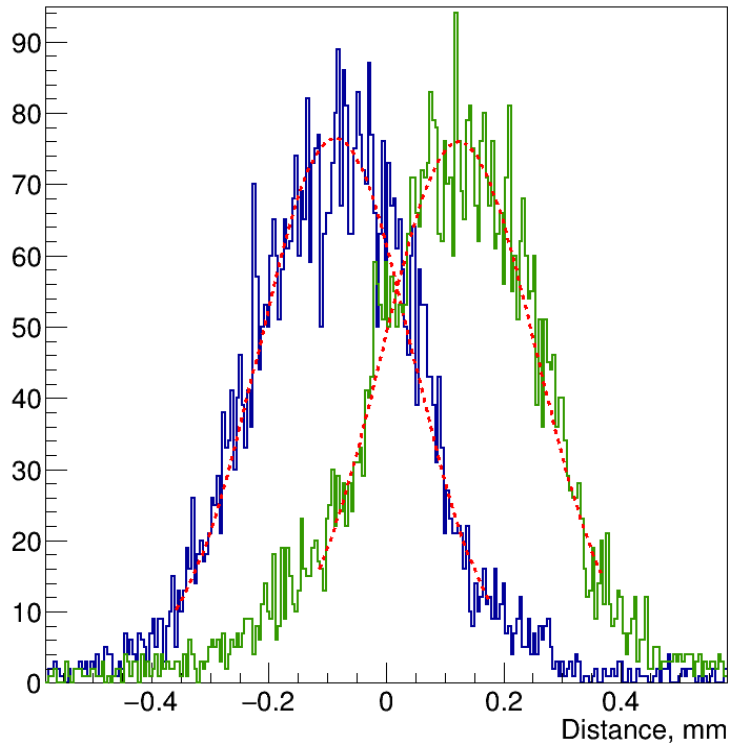

Figure 17. Projections of the slit image obtained from the dataset with *W* value above 1.39 (blue peak on the left) and below this value (green peak on the right). The red dashed curves show Gaussian fits of the peaks. The separation between the peaks is 0.21 mm and the width of both peaks (FWHM) is 0.31 mm.

## 3.3 Reconstruction with one strip disabled

This section is dedicated to the analysis of the reconstruction quality for the situations when the signals recorded with one of the strips have to be disregarded (for example, due to lost contact, strongly drifted gain of the preamplifier leading to saturation or a DAQ-related problem).

For the centroid reconstruction loss of one channel is critical. The reconstructed event positions are pulled away from the true position as demonstrated in figure 18 (left). The images of a diagonally-oriented 0.4 mm wide slit were obtained disregarding the signals of the strip situated at Y of -7 mm. For the statistical reconstruction the loss of one channel is much more tolerable: figure 18 (right) demonstrates that the reconstruction images do not show noticeable distortions.

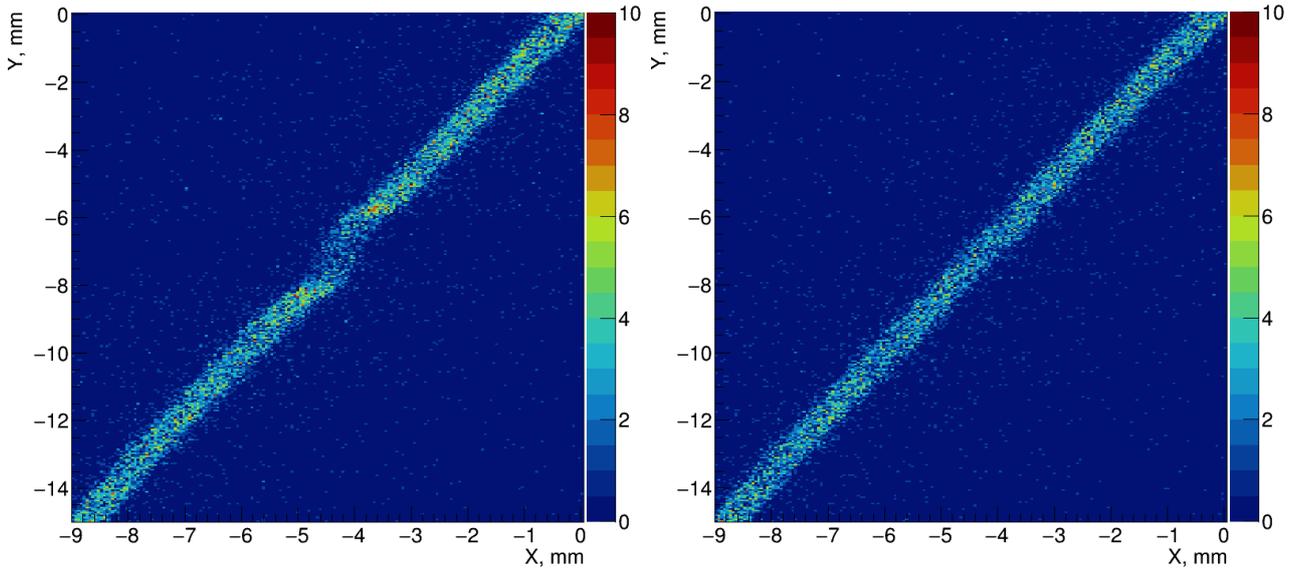

Figure 18. Heatmaps of the reconstructed event density (number of events per bin) for a dataset recorded with a 0.4 mm wide slit oriented diagonally using the centroid (left) and statistical (right) reconstruction, performed disregarding the signals of the horizontal pick-up strip situated at Y = 7 mm. Bin size is 0.05 mm in both directions.

Figure 19 shows the results of the Gaussian fit for the datasets recorded with a vertical slit at X positions ranging from 1 to 5 mm with the step of 0.1 mm (the same data were used in figure 14) and reconstructed with the centroid and statistical methods disregarding the signals of the strip at X of 3 mm. The distributions of the mean of the Gaussian fit of the profiles demonstrate strong non-linearity of the centroid reconstruction in these conditions, while linearity of the statistical method remains essentially the same as in the case when signals from all strips are considered (figure 14).

The apparent resolution for the centroid reconstruction is strongly affected by the local non-linearity and thus should not be analyzed before application of the linearity correction. Such a correction at a given X position can be performed based on the tangent of the local linear fit of the top-left graph of figure 18 at the same X position, which characterizes the "compression" or "expansion" of the local reconstructed space: in the region of compression, the reconstructed image appears smaller than the true one, and the reconstructed features are situated closer compared to the true distance between them. The opposite behavior is observed in the region of expansion. A correction factor of 0.49 obtained at X = 3 mm (the angle of ~65 degrees) suggest that the resolution for the centroid method at this position is about 0.31 mm. Note that the same resolution is obtained for the statistical method. Similarly, at X = 4 mm the correction factor is 1.3 (~37 degrees), which results in a resolution of about 0.35 mm, which is also consistent with the one given by the statistical method at this position.

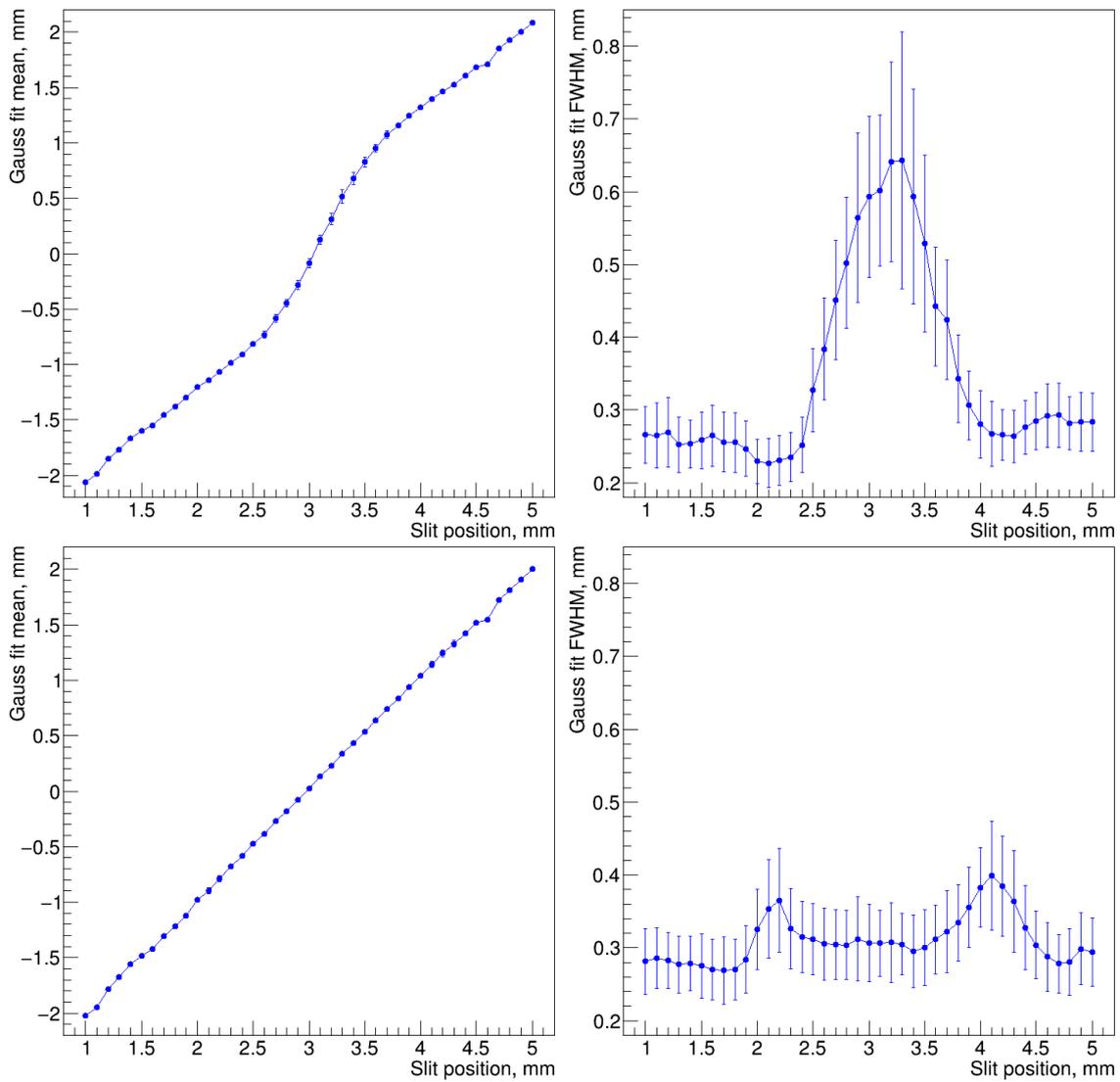

Figure 19. Mean (left) and FWHM (right) of the Gaussian fit of the profile for 41 datasets recorded with a vertical slit at X positions ranging from 1 to 5 mm with the step of 0.1 mm. The top and bottom rows give centroid and statistical reconstruction results, respectively. Both reconstructions were performed disregarding the signals of the strip at X = 3 mm. The error bars show the uncertainties of the fit. For the mean, the uncertainties are smaller than the size of the markers.

The reason for the good tolerance of the statistical reconstruction to a channel loss is explained by the capability of the fit to give a quite precise estimate for the profile center using signals from the remaining strips. Figure 20 shows examples of the fit of the experimental strip signals performed with all the strips (left hand side), as well as for the cases when the strip with the strongest (middle) and the second-strongest signal (right hand side) are disabled. The fit results are very similar in all three cases. Note that for weaker events the fit can be less accurate when a strip is disabled.

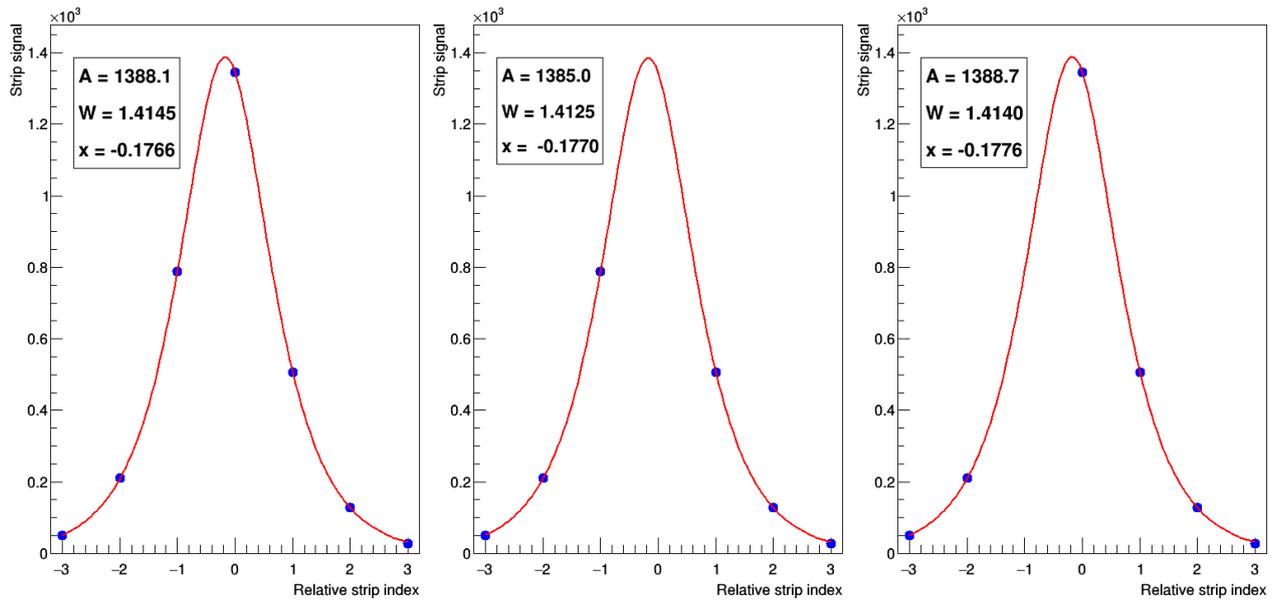

Figure 20. Red curves show fits of the experimental signals (blue dots) performed considering all strips (left hand side), as well as for the cases when the strip with the strongest (middle) and the second-strongest signal (right hand side) are disabled. The values of the fit parameters are given in the text boxes, showing a very small difference in the value of x.

## 3.4 Relative strength of the read-out channels

Before shipping the detector to HZB, all electronic channels were tested using a pulse generator confirming that their relative gains are within 2%. However, the distortion that appears in all reconstructed images at Y of about 7 mm can be explained if we assume that the gain of one channel has changed during the detector transportation. This conclusion is based on two facts: the width of the distorted area is comparable with the width of the strip response function, and the contribution to $\chi^2$ from the strip situated at Y of 7 mm is significantly stronger than that of the other strips.

This hypothesis can be tested benefiting from the capability of the statistical reconstruction to provide a good fit of the event signals even when one of the channels is disregarded. The relative strength of a particular electronic channel in respect to several neighboring ones can be evaluated by the following procedure. For each event in a flood-field dataset, using equation (1) make a fit of the observed signals disregarding that channel. Compute the ratio of the experimental signal in the disregarded channel to the value given by the equation (1) based on the fit. The relative strength is then given by the average value of this ratio over all events of the dataset. The result of this procedure performed for each strip in Y direction is shown in figure 21.

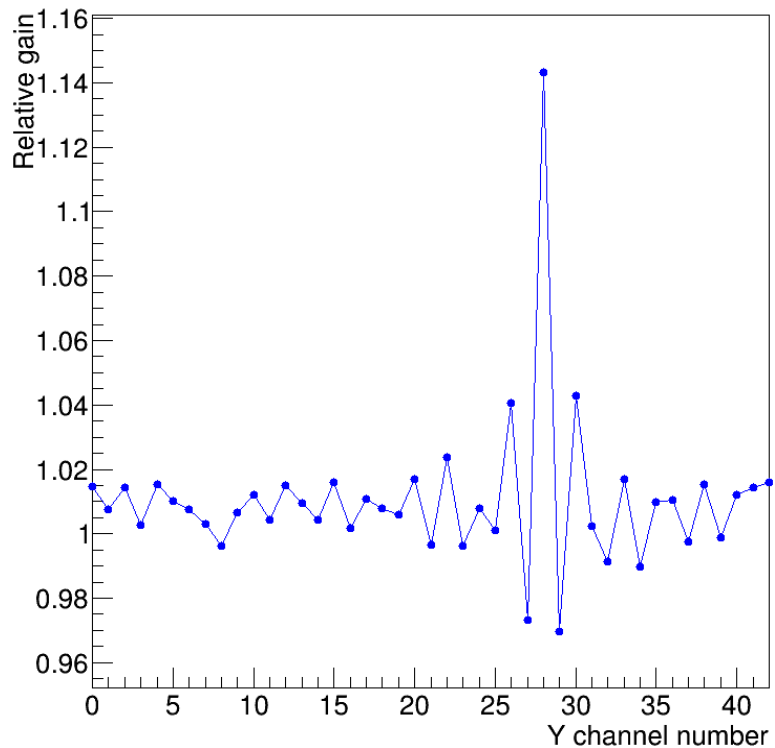

Figure 21. Relative strength (in respect to the neighbors) of the channels in Y direction. The Y position of each strip (in mm) is given by the channel number minus 21.

The results show that the relative strength for most of the channels in Y direction are different by less than 2%, which is consistent with the calibration mentioned at the beginning of this section. However, the channel #28 (Y = 7 mm) is ~15% stronger than the average value. Note the ripple-like pattern in the values around the channel #28: two neighboring ones (#27 and #29) have smaller-than-average strength while the second-neighbor channels (#26 and #30) have larger-than-average strength. An analysis of the distortions of the fit in the condition when all channels have the same strength with exception of one channel which is ~15% stronger than the rest shows that this is an expected pattern.

Assuming that the relative strength of the channel #28 is 1.15, statistical reconstruction of the flood irradiation dataset indeed gives somewhat better results in the area close to Y = 7 mm (compare the images in figure 22).

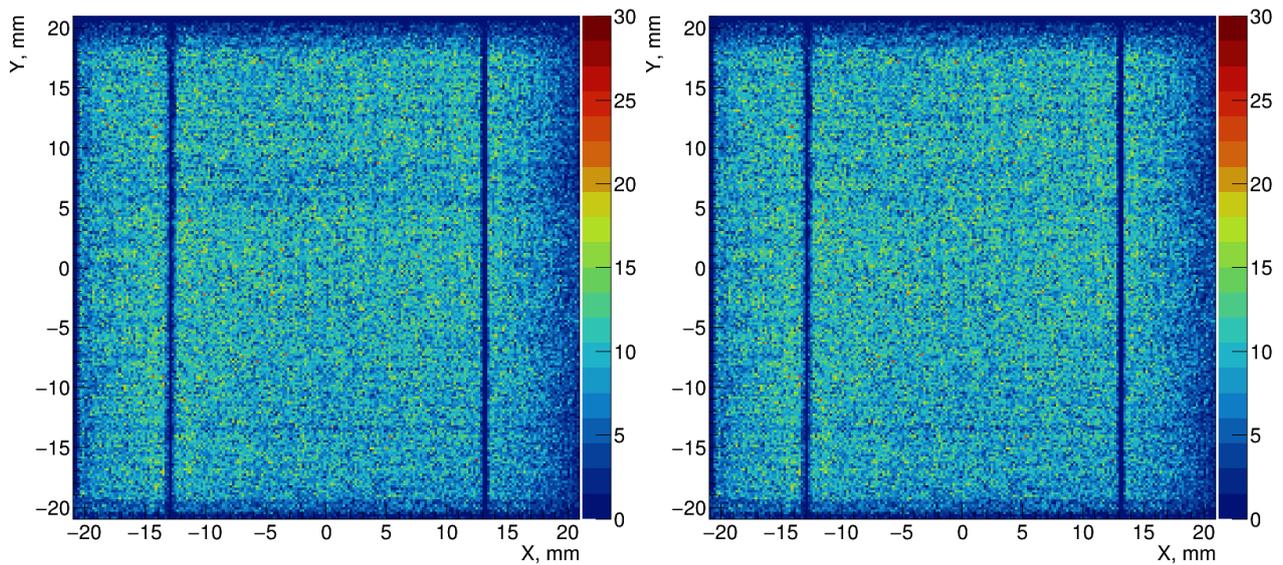

Figure 22. Heatmaps of the reconstructed event density (number of events per bin) for the flood irradiation dataset. The reconstruction is performed with the statistical method assuming that all channels have the same relative strength (left) and that the channel #28 has a relative strength 15% higher than all other channels (right).

The described procedure is effective only when just one strip has unknown relative strength. However, it might be possible to construct an iterative procedure which calculates relative gains for all the channels by evaluating the relative strength of one channel at a time. Development and validation of such procedure requires a dedicated study.

## 4. Discussion

The results of this study show that both reconstruction methods can provide similar image quality. With the exception of the detector periphery, where the statistical reconstruction performs better (see figure 7), the uniformity of the reconstructed images is essentially the same for both methods. The linearity of the images is very similar: compare, for instance, the images forming the "FRM II" letters in figure 11. Also, in the conditions of this study, both methods provide the same spatial resolution.

However, the statistical reconstruction method has several advantages. It offers a broader set of event filtering techniques, which can be very efficient in suppression of "bad" events that deviate from the response model defined by the strip response functions. It is also easier to define the filter ranges from analysis of the flood irradiation datasets. The mathematical model of the response allows to develop an efficient early warning system, based, for example, on analysis of $\chi^2$ (see figure 9). Such a system would be able to generate an alarm if during the detector operation some of the critical parameters (e.g., the electronic gain of one of the channels) deviates from the acceptable range. The statistical reconstruction has a certain level of in-build redundancy which allows, for example, extract position information without a drastic loss in the accuracy if signals from one of the pick-up strips (or several sufficiently-separated ones) have to be disregarded. Finally, this method seems to be less affected by the electronic noise, and thus can, potentially, offer better ultimate spatial resolution. This statement, however, requires a confirmation based on a dedicated study aimed at precise characterization of all factors contribution to the spatial resolution.

The price of the application the statistical method is the computation complexity. For the implementation developed in this study (not optimized for speed), the statistical reconstruction took one order of magnitude longer time compared to the centroid reconstruction. A possible solution is to implement the statistical reconstruction on a graphics processing unit (GPU) using, for example, the contracting grids algorithm (see, e.g., implementation described in [17]). We expect that in this case the processing times of millions events per second can be reached.

This study also demonstrates that for detectors based on double gap RPCs with interconnected pick-up strips at the same X (and, similarly, Y) coordinate, the readout schemes which do not allow to distinguish events localized in the different gas-gaps of the same double-gap RPCs should be avoided. The inability to identify the triggered gas-gap leads to the blurring of the reconstructed images if there is a non-negligible shift (in comparison with the position resolution) between the corresponding strip positions or if the neutron beam has a non-negligible deviation of the angle of incidence from the normal direction to the RPC plane. To provide the possibility to identify the triggered gas-gap, in the next detector prototype we plan to rotate the pick-up strip arrays on one side of the double gap RPC by 90 degrees. In this case the inner array (see figure 1) will have horizontal strip orientation on one side and vertical on the other, thus allowing to identify the gas-gap using the fact that the sum signal induced in the inner array has ~30% stronger amplitude compared to that of the outer array.

## 5. Conclusions

A new statistical-based approach for position reconstruction for RPC-based detectors was developed and experimentally validated. The results of this study show that the centroid and the statistical methods result in similar image quality and spatial resolution. However, the statistical method allows to improve the image quality at the periphery of the detector, offers more flexible event filtering, is less affected by the electronic noise and allows to develop automatic quality monitoring procedures for detection of situations when a change in the detector operation conditions starts to affect the image quality. It was also demonstrated that the readout scheme of $^{10}$B-RPC detectors should allow to distinguish events localized in the different gas-gaps of the double-gap RPCs. It is needed to avoid blurring of the reconstructed images due to misalignment of the signal pick-up strips and the parallax effect for neutron beams with non-orthogonal incidence.

## Acknowledgments

This work was supported by Portuguese national funds OE and FCT-Portugal (grant CERN/FIS-INS/0009/2019). K. Roemer acknowledges the support from the European Fund for Regional Development and the Program for R&D of the Sächsische Aufbaubank under the code WIDDER-100325989. The authors also acknowledge the Helmholtz-Zentrum Berlin for providing the beam time at the V17 Detector Test Station.